\documentclass[11pt,usletter]{article}
\usepackage{jheppub}
\pdfoutput=1

\usepackage{graphicx}
\usepackage{hyperref}
\usepackage{color}

\def\ba{\begin{align}}
\def\ea{\end{align}}
\def\be{\begin{equation}}
\def\ee{\end{equation}}
\def\bea{\begin{eqnarray}}
\def\eea{\end{eqnarray}}

\begin{document}

\title{Tracing Primordial Black Holes in Nonsingular Bouncing Cosmology}

\author[a]{Jie-Wen Chen,}
\author[a,b]{Junyu Liu,}
\author[a,c]{Hao-Lan Xu,}
\author[a,*]{Yi-Fu Cai}

\affiliation[a]{CAS Key Laboratory for Researches in Galaxies and Cosmology, Department of Astronomy, University of Science and Technology of China, Hefei, Anhui 230026, China}
\affiliation[b]{Department of Physics, California Institute of Technology, Pasadena, California 91125, USA}
\affiliation[c]{Institut d'Astrophysique de Paris, UMR 7095-CNRS, Universit\'{e} Pierre et Marie Curie, 98 bis boulevard Arago, 75014 Paris, France}

\emailAdd{chjw@mail.ustc.edu.cn}
\emailAdd{junyu@mail.ustc.edu.cn}
\emailAdd{xhl1995@mail.ustc.edu.cn}
\emailAdd{yifucai@ustc.edu.cn}

\date{\today}

\abstract{
We in this paper investigate the formation and evolution of primordial black holes (PBHs) in nonsingular bouncing cosmologies.
We discuss the formation of PBH in the contracting phase and calculate the PBH abundance as a function of the sound speed and Hubble parameter.
Afterwards, by taking into account the subsequent PBH evolution during the bouncing phase, we derive the density of PBHs and their Hawking radiation.
Our analysis shows that nonsingular bounce models can be constrained from the backreaction of PBHs.
}


\maketitle
\flushbottom

\section{Introduction}

The matter bounce scenario \cite{Brandenberger:2012zb, Wands:1998yp, Finelli:2001sr} is one type of nonsingular bounce cosmology \cite{Brandenberger:2009jq, Brandenberger:2016vhg, Novello:2008ra, Lehners:2008vx, Cai:2014bea, Battefeld:2014uga}, which is often viewed as an important alternative to the standard inflationary paradigm \cite{Guth:1980zm, Starobinsky:1980te, Sato:1980yn, Fang:1980wi}. By suggesting that the universe was initially in a contracting phase dominated by dust-like fluid (with a vanishing equation-of-state parameter $w=0$), then experienced a phase of nonsingular bounce, and afterwards entered a regular phase of thermal expansion. The matter bounce cosmology can solve the horizon problem as successfully as inflation and match with the observed hot big bang history smoothly. Based on primordial fluctuations generated during matter contracting and their evolution through the nonsingular bounce, one can obtain a scale invariant power spectrum of cosmological perturbations. Unlike inflation, the matter bounce scenario does not need a strong constrain on the flatness of the potential of the primordial scalar field that drives the evolution of the background spacetime \cite{Adams:1990pn, Copeland:1994vg}. Also, this scenario can avoid the initial singularity problem and the trans-Planckian problem, which exists in inflationary and hot big bang cosmologies \cite{Borde:1993xh, martin2001trans}.

The aforementioned scenario has been extensively studied in the literature, such as the quintom bounce~\cite{Cai:2007qw, Cai:2007zv}, the Lee-Wick bounce \cite{Cai:2008qw}, the Horava-Lifshitz gravity bounce \cite{Brandenberger:2009yt, Cai:2009in, Gao:2009wn}, the $f(T)$ teleparallel bounce \cite{Cai:2011tc, deHaro:2012zt, Cai:2015emx}, the ghost condensate bounce \cite{Lin:2010pf}, the Galileon bounce \cite{Qiu:2011cy, Easson:2011zy}, the matter-ekpyrotic bounce \cite{Cai:2012va, Cai:2013kja, Cai:2014zga}, the fermionic bounce \cite{Alexander:2014eva, Alexander:2014uaa}, etc.~(see, e.g. Refs. \cite{Brandenberger:2010dk, Brandenberger:2012zb} for recent reviews). In general, it was demonstrated that on length scales larger than the time scale of the bouncing phase, both the amplitude and the shape of the power spectrum of primordial curvature perturbations can remain unchanged through the bouncing point due to a no-go theorem \cite{Quintin:2015rta, Battarra:2014tga}. A challenge that the matter bounce cosmology has to address is how to obtain a slightly red tilt on the nearly scale invariant primordial power spectrum. To address this issue, a generalized matter bounce scenario, which is dubbed as the $\Lambda$-Cold-Dark-Matter ($\Lambda$CDM) bounce, was proposed in \cite{Cai:2014jla} and predicted an observational signature of a positive running of the scalar spectral index \cite{Cai:2015vzv, Cai:2016hea}.

As a candidate describing the very early universe, the matter bounce scenario is expected to be consistent with current cosmological observations and to be distinguishable from the experimental predictions of cosmic inflation as well as other paradigms \cite{constraint, Cai:2014bea}. Meanwhile, a possible probe of primordial black holes (PBHs) may offer a promising observational approach to distinguish various paradigms of the very early universe \cite{Carr:1975qj, Carr:2009jm}. PBHs could form at very early times of the universe, where a large amplitude of density perturbations would have obtained. Correspondingly, the formation process and the abundance of PBHs strongly depend on those early universe models, in which fluctuations of matter fields are responsible for such large amplitudes of density perturbations \cite{Carr:1974nx}.

In the literature, most of attentions were paid on the computation of PBH predictions from the inflationary paradigm (for instance see \cite{sub-Hubble,sub-Hubble2, saperate, threshold, Josan}), while so far, only a few works addressed the PBH formation in a bouncing scenario\cite{carr2011persistence, carr2016primordial}.
Furthermore, those studies of PBHs in a bouncing scenario have not yet been discussed in detail,
for specific cosmological paradigms or been applied to falsify various early universe cosmologies, especially the matter bounce scenario.
In the context of matter bounce cosmology, there are several differences on the computation of the PBH abundance comparing with that in an expanding universe. First, comparing with inflation where the primordial fluctuations become frozen at the moment of the Hubble exit, those primordial fluctuations on matter fields in bounce cosmology would continue to increase after the Hubble exit during the contracting phase until the universe arrive at the bouncing phase \cite{Cai:2008qw, Brandenberger:2016vhg}, and the contracting phase would yield a different initial condition for the PBH formation and evolution.
Second, once these PBHs have formed, the contraction of spacetime could also compress and enlarge the primordial matter density, thus change the PBH horizon radius which then can lead to effects on their evolution.

In this paper, we perform a detailed survey on the PBH formation and evolution in the background of the matter bounce cosmology. In Section \ref{model}, we briefly introduce the matter bounce scenario and describe the formation of the power spectrum of primordial curvature perturbation in an almost model-independent framework. In Section \ref{formation}, a physical picture of the PBH formation in the contracting background is presented. After a process of detailed calculations, the threshold for forming PBHs and the corresponding mass fraction are provided. In Section \ref{evolve}, we discuss the evolution of PBHs in the bouncing phase by taking into account the effects arisen from the contraction of the background and the Hawking radiation. In Section \ref{conclusion}, we summarize our results and discuss on some outlook of the PBH physics within the nonsingular bouncing cosmology.

\section{Nonsingular bounce}\label{model}

Nonsingular bounce can be achieved in various theoretical models, namely, to modify the gravitational sector beyond Einstein, to utilize matter fields violating the Null Energy Condition (NEC), or in the background of non-flat geometries (see e.g. \cite{Martin:2003sf, Solomons:2001ef}). It is interesting to notice that, in general, on length scales larger than the time scale of the nonsingular bouncing phase, primordial cosmological perturbations remain almost unchanged throughout the bounce \cite{Quintin:2015rta, Battarra:2014tga}. In this regard, one expects that the effective field theory approach should be efficient to describe the information of a nonsingular bounce model at background and perturbation level. Recently, it was found in \cite{Cai:2012va} that a nonsingular bounce model can be achieved under the help of scalar field with a Horndeski-type, non-standard kinetic term and a negative exponential potential. Within this model construction, the matter contracting phase can be obtained directly by including the dust-like fluid or involving a second matter field \cite{Cai:2013kja}.
Note that, in the present study we assume that the effective field approach of bouncing cosmology is valid through the whole evolution without modifications to General Relativity.

\subsection{The model}

It turns out that, under the description of the effective field theory approach, the background dynamics of the nonsingular bouncing cosmology can be roughly separated into three phases: the matter-dominated contraction, the non-singular bounce, and the thermal expansion. We consider a simple model starting with a matter contracting phase ($t<t_{-}$) from an initial time $t_{\text{initial}}\rightarrow-\infty$, and then entering into a nonsingular bouncing phase at $t_{-}$, which lasts till $t_{+}$. After the bounce ends at $t_{+}$, the universe begins the hot big bang expansion, which is in accordance to the current observations.

The evolution of the matter bounce cosmology in each stage can be approximately described as follows.\\
(i) In the matter contracting phase, the scale factor of the universe shrinks as
\be
 a(t)=a_{-} \left(\frac{t-\tilde{t}_{-}}{t_{-}-\tilde{t}_{-}}\right)^{2/3}~,
\ee
where $a_{-}$ is the scale factor at time $t_{-}$, and $\tilde{t}_{-}$ is related to the Hubble parameter at $t_{-}$ via the relation $t_{-}-\tilde{t}_{-}=\frac{2}{3H_{-}}$. The equation-of-state parameter during this phase is $w=0$, which can be realized in many ways, such as by cold dust, by massive field or by the gravity sector involving non-minimal couplings. We parameterize these different mechanisms by introducing the sound speed $c_s$, which can affect the propagation of primordial perturbations in the gradient terms.\\
(ii) In the nonsingular bouncing phase, the scale factor of the universe can be approximately described as \cite{Cai:2012va}
\be
\label{bouncing}
a(t)=a_\text B e^{\frac{\Upsilon t^2}{2}}~,
\ee
where the coefficient $a_\text B$ is the scale factor exactly at the bouncing point, and from Eq. (\ref{bouncing}) one obtains $a_{-}=a_\text B \exp [\Upsilon t_{-}^2/{2}] $. $\Upsilon$ is a model parameter describing the slope of Hubble parameter to time during the bouncing phase, as:
\be
\label{bouncingH}
 H(t)=\Upsilon t~.
\ee
It can be seen that $t=0$ corresponds to the bouncing point when the universe stops the contraction and starts the expansion.
Thus, it can be found that $t_{-}=H_{-}/\Upsilon$, and the value of $H$ vanishes at $t=0$ which is at the bouncing point. Inserting Eqs. (\ref{bouncing}) and (\ref{bouncingH}) into the Friedmann equation,
one can see that the null energy condition $\rho + p > 0$ is violated around the bouncing point. It is the negative pressure that avoids the singularity and drives the universe to evolve from a contracting phase to an expanding phase.\\
(iii) In the era of radiation-dominated expansion, we have
\be
 a(t)=a_{+} \left(\frac{t-\tilde{t}_{+}}{t_{+}-\tilde{t}_{+}}\right)^{1/2}~,
\ee
where $t_{+}=H_{+}/\Upsilon$, $t_{+}-\tilde{t}_{+}=\frac{1}{2H_{+}}$ and $a_{+}=a_\text B e^{\frac{\Upsilon t_{+}^2}{2}}$. In present analysis we have adopted the assumption that the heating process happens instantly after the bounce (see \cite{Quintin:2014oea, deHaro:2015wda} for relevant analyses).

\begin{figure}
\begin{center}
\includegraphics[width=0.7\textwidth]{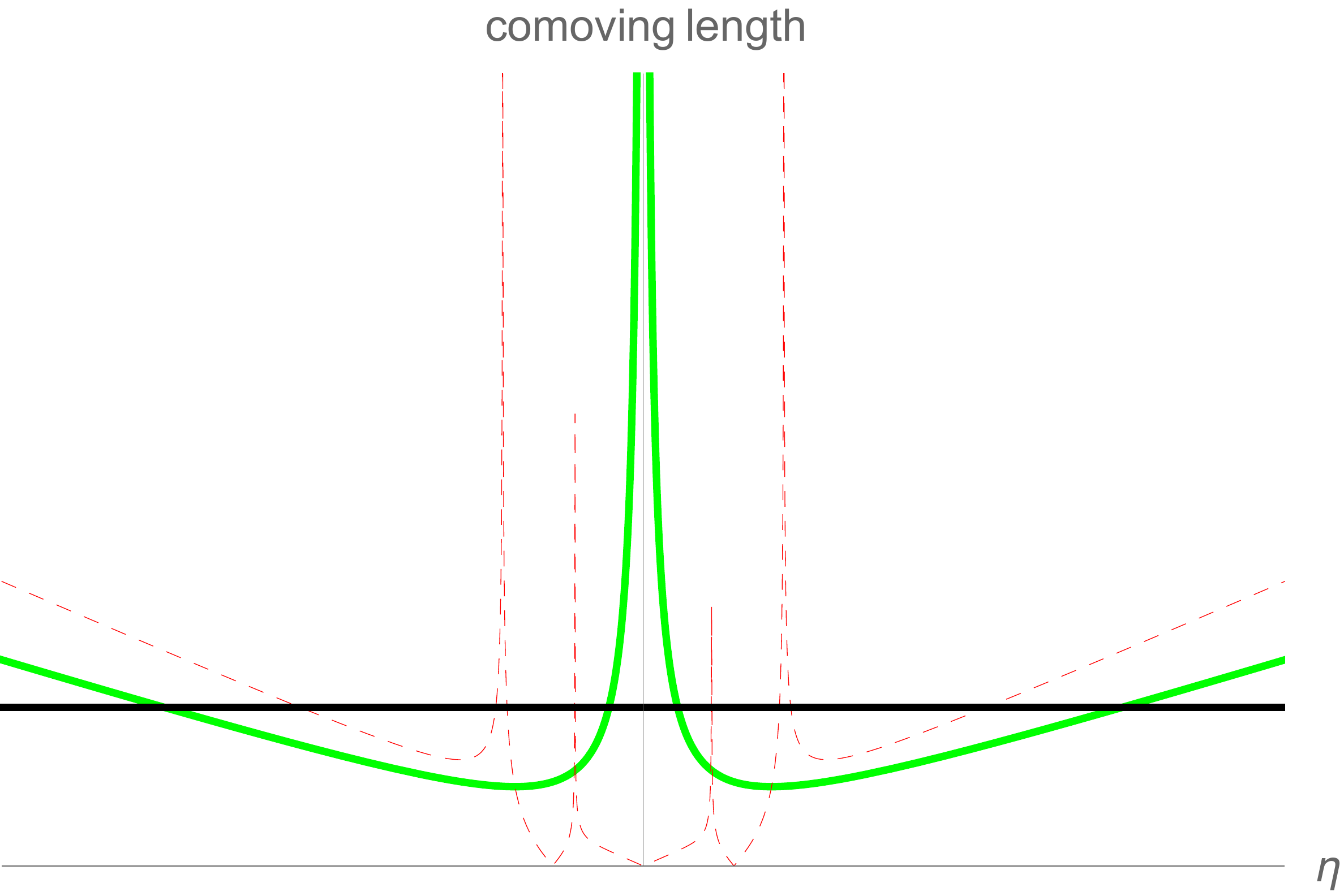}
\end{center}
\caption{
\label{comhubble}
Sketches of the evolutions of the comoving Hubble radius (green curve) $|\mathcal H|^{-1}=|a H|^{-1}$
and an effective length scale from the coefficient $|z''/z|^{-1/2}$ (red dashed) where $z=\frac{a \sqrt{\rho+p}}{H}$,
in the nonsingular bounce cosmology. The black curve is the comoving length for wavenumber $k$.
The horizontal axis corresponds to the comoving time which is defined by $\eta\equiv \int \frac{dt}{a}$.
}
\end{figure}

From such parameterizations, the matter bounce cosmology can be approximately described by model parameters $H_{-}$, $H_{+}$, $\Upsilon$, and also $c_s$ if perturbations are taken into account. In Fig.~\ref{comhubble} we depict the evolution of comoving Hubble length $|\mathcal H^{-1}|=|a H|^{-1}$, and one can read that one Fourier mode of cosmological perturbation in the matter bounce cosmology could exit the Hubble radius during the contracting phase, and then enter and re-exit the Hubble radius during the bouncing phase, and eventually re-enter the Hubble radius again in the classical Big Bang era. Note that the change of the Hubble radius in the vicinity of the bouncing point can be very large. In the literature, observational constraints upon the bounce cosmology can be derived from various cosmological experiments such as the cosmic microwave background \cite{constraint} and primordial magnetic fields \cite{Qian:2016lbf}. In this work we will provide the independent constraints on the model parameters from PBHs.

\subsection{Curvature perturbation during matter contracting phase} \label{formation}

During the matter contracting phase, equation of motion for the curvature perturbation can be expressed as
\be
\label{EoM}
v_k''+(c_s^2 k^2-\frac{z''}{z})v_k=0~,
\ee
where $v_k=z \mathcal R_k$ is the Mukhanov-Sasaki variable with $\mathcal R_k$ being the comoving curvature perturbation, $z=\frac{a \sqrt{\rho+p}}{c_s H}$ relies on the detailed evolution of the background dynamics, and the prime represents the derivative with respect to the comoving time $\eta$. Assuming that primordial perturbations originated from vacuum fluctuations at initial times\footnote{Note that the vacuum state is not necessarily the only choice of the initial condition for primordial curvature perturbations, namely, they may also arise from fluctuations of a thermal \textbf{ensemble} \cite{Cai:2008qw, Cai:2009rd}.},
one can derive the solution of Eq. \eqref{EoM}
\be
\label{MS}
v_k=\frac{\sqrt{\pi(-\eta)}}{2} H_{3/2}^{(1)}[c_s k (-\eta)]~,
\ee
where $H_{3/2}^{(1)}$ is the $\frac{3}{2}$-order Hankel function of the first kind.
From Eq. \eqref{MS}, the power spectrum of $\mathcal R_k$ is then given by
\be
\Delta^2_{\mathcal R}\equiv\frac{k^3}{2\pi^2}\left|\frac{v_k}{z}\right|^2=\frac{c_s^2 k^3 (-\eta)}{24\pi M_\text p^2 a^2}\left|H_{3/2}^{(1)}(-c_s k \eta)\right|^2~,
\ee
during the matter contracting phase. At large scales $c_s k\ll |\mathcal H|$, $\Delta^2_{\mathcal R}\simeq \frac{a_{-}^3 H_{-}^2}{48\pi^2 c_s M_\text p^2 a^3}$ is almost scale independent, and $\mathcal R_k \simeq \sqrt{\frac{a_{-}^3 H_{-}^2}{24 k^3 c_s M_\text p^2}}a^{-3/2}$.
This result is different from that in an expanding spacetime, in which $\mathcal R_k$ is time independent at large scales.
Moreover, at small scales $c_s k \gg |\mathcal H|$, curvature perturbation is of quantum fluctuation with $\Delta^2_{\mathcal R}\simeq \frac{a_{-}^3 H_{-}^2}{12\pi^2 c_s M_\text p^2 a^3}\left(\frac{c_s k}{\mathcal H}\right)^2$.

\section{PBH formation during matter contracting phase}\label{formation}

PBHs originate from the collapsed over-dense regions seeded by cosmological perturbations in early universe.
When an over-dense region starts to form a black hole (BH), its radius $R$ is expected to be larger than the Jeans radius $R_\text J\equiv\frac{c_s}{2}\sqrt{\frac{\pi}{\bar \rho}}=\sqrt{\frac{2}{3}}\pi c_s  |H^{-1}|$
and smaller than the Hubble length $|H^{-1}|$ \cite{Carr:1975qj, Carr:1974nx, threshold, Gao:2011tq, Quintin:2016qro}, which implies,
\be
\label{RRR}
\sqrt{\frac{2}{3}}\pi c_s  |H^{-1}| \leq R \leq |H^{-1}|~.
\ee
For scales smaller than $R_\text J$, the pressure gradient force would prevent the collapsing of the over-dense region.
It is clear that $R_\text J \geq |H^{-1}|$ when $c_s \geq 0.39$, and hence, there would be no PBH in this case.

We note that the PBH formation in the contracting phase is different from that in a purely expanding universe.
In the expanding universe, the PBH formation would begin only when the size of a region reenters the Hubble scale.
However, in the contracting phase, when the size of the region exits the Jeans scale, the PBH formation would have started if this region is dense enough.
Thus, we would like to focus on the fluctuations within the Jeans scale instead of the Hubble scale in the contracting universe.

\subsection{PBH mass fraction}

The background dust-like fluid inside the over-dense regions may collapse into PBHs, and
the abundance of PBHs is depicted by the mass fraction, which in the matter contracting phase is
\be
\label{betadef}
\beta=\frac{\rho_{\text {PBH}}}{\rho_{\text {PBH}}+\rho_{\text {bg}}}~,
\ee
where $\rho_{\text {PBH}}$ denotes the density of PBHs and $\rho_{\text {bg}}$ the density of the remnant dust-like fluid.
In general, PBHs and remnant dust-like fluid together drive the evolution of background $\rho_{\text {PBH}}+\rho_{\text {bg}}=3M_\text p^2 H^2$.
Note that this relation only holds when the subsequent evolutions, including the accretion and Hawking radiation, of the formed PBHs are not considered. 

The mass fraction can be obtained from the Press-Schechter theory \cite{Press:1973iz}(see \cite{Carr:1974nx, Carr:1975qj, threshold} also):
\be
\label{PS}
\beta(t)=\int_{\delta_\text c}^{\delta_\text m}\frac{2}{\sqrt{2\pi\sigma}} \exp(-\frac{\delta^2}{2\sigma^2})\text d\delta
={\rm erfc} \left(\frac{\delta_\text c}{\sqrt{2}\sigma(t)}\right) {-{\rm erfc} \left(\frac{\delta_\text m}{\sqrt{2}\sigma(t)}\right)}~,
\ee
where $\delta \equiv \frac{\delta \rho}{\bar \rho}$ is the fractional density fluctuations, $\sigma\equiv \sqrt{\langle\delta^2\rangle}$ is the mean mass fluctuation at the Jeans scale, which can be derived from the power spectrum of curvature perturbation $\Delta^2_{\mathcal R}$,
$\delta_\text c$ is the threshold of the PBH formation, and the upper limit $\delta_\text m$ ensures that the PBHs are no larger than the Hubble scale  \cite{Carr:1974nx, Carr:1975qj, threshold}.
Note that Eq. (\ref{PS}) is based on the assumption that $\delta$ obeys the Gaussian distribution ${N}[0,\sigma]$.


The value of threshold $\delta_\text c$ is determined as follows. Considering a spherical over-dense region with a radius $R$,
the space inside the region satisfies  Friedmann equation  \cite{gunn1972infall, peebles1967gravitational}
\be
\label{FRW}
\left(H+\delta H\right)^2=H^2\left(1 + \delta \right) - \frac{\delta K}{a^2}=H^2(1+\tilde{\delta})~,
\ee
where $H$ is the Hubble parameter of the background, $a$ is the scale factor inside the over-dense region, $\delta H$ and $\delta K$ are the perturbed Hubble parameter and curvature respectively, and $\tilde{\delta}\equiv \delta-\frac{\delta K}{a^2 H^2}$. 
It is noticed that $H <0$ for a contracting phase, and $\delta H<0$ in the over dense region. The outer region ($>R$) is thought to be unperturbed for simplicity. We assume that when the region collapses to a BH, its surface has an additional physical speed $v=1$ with respect to the conformally static background, i.e. collapsing at the speed of light. In the comoving slicing, it is written as
\be
\label{BH forming}
\delta H \cdot R = -1~.
\ee
Inserting Eq. (\ref{FRW}) into the above equation, one obtains the threshold for an arbitrary scale $R$
\be
\label{threshold}
\tilde \delta_\text c=\delta_\text c - \frac{\delta K_\text c}{a^2 H^2}=\frac{1}{H^2 R^2}+\frac{2}{|H| R}~.
\ee
On the Jeans scale $R_\text J$, one has $\tilde \delta_\text c=\frac{\sqrt{6}}{\pi c_s}+\frac{3}{2\pi^2 c_s^2}$.

The mass of BH with a radius $R$, in the contracting background, can be determined by
\be
\label{M(R)}
M(R)=\frac{4\pi}{3}R^3\bar\rho \left(1+\tilde \delta_\text c \right)=\frac{R}{2G}(1+|H|R)^2~.
\ee
Note that, both the density and curvature fluctuations could contribute to the BH mass, as shown in \cite{threshold}.
We mention that the above description of BH is a rough estimate,
and hence, Eq. (\ref{M(R)}) slightly differs from the Misner-Sharp mass $M=\frac{R}{2G}$.
The accurate description of the BH in the matter contracting phase should be based on the Tolman-Bondi-Lemaitre metric \cite{TBL1,TBL2,TBL3}, which will be addressed in our follow-up study. In present analysis, however, we note that at small scale $R\ll |H^{-1}|$ the above model naturally recovers the solution of a Schwarzschild BH $M=\frac{R}{2G}$ in a flat spacetime, which indicates that this estimate remains reliable.

To derive the values of $\delta_\text c$ and $\sigma$, one needs to know the relations among the variables $\delta$, $\delta K$ and $\mathcal R_k$. It is convenient to take the comoving gauge, in which
\be
\label{k and zeta}
H^2 \delta= \frac{2}{3}\nabla^2 \Psi(x,t)~, \ \ \ \ \ \ \
\frac{\delta K}{a^2}= -\frac{2}{3} \nabla^2 \mathcal R(x,t)~,
\ee
where $\Psi(x,t)$ is the Bardeen potential and $\mathcal R(x,t)$ is the curvature perturbation\cite{curvature, Liddle-Lyth, sub-Hubble}. The relation between their Fourier components $\Psi_k$ and $\mathcal R_k$ is given by \cite{Liddle-Lyth, sub-Hubble}
\be
 -(1+w)\mathcal R_k=\frac{5+3w}{3}\Psi_k +\frac{2\dot \Psi_k}{3H}~,
\ee
where the dot denotes the derivative with respect to the cosmic time $t$. The solution during the matter contracting era $\mathcal R_k \propto a^{-3/2}$ is
\be
\Psi_k=-\frac{3}{2}\mathcal R_k~,
\ee
differing from $\Psi_k=-\frac{3}{5}\mathcal R_k$ in the matter dominant expanding era. One also obtains
\be
\label{delta spectrum}
\delta_k=\frac{k^2}{a^2 H^2} \mathcal R_k~,
\ee
where $\delta_k$ is the Fourier component of the density perturbation $\delta$. From (\ref{threshold}) and (\ref{k and zeta}), one also has $\delta=\frac{3\delta K}{2a^2 H^2}=3 \tilde \delta$.
Therefore, the threshold is $\delta_\text c= \frac{3 \sqrt{6}}{\pi c_s}+\frac{9}{2\pi^2 c_s^2}$.

The value of the upper limit $\delta_\text m$ for the Jeans scale is fixed as follows.
If $\delta=\delta_\text m$, the fluid inside the Hubble radius would form a BH
\footnote{In this case, the PBH is formed by an over-dense core with radius $R_J$, and a shell of the background fluid within the radius from $R_J$ to $|H^{-1}|$.},
which would have
\be
\frac{4\pi}{3}R_\text J^3\bar\rho \left(1+\tilde \delta_\text m \right)+\frac{4\pi}{3}\bar\rho(|H^{-3}|-R_\text J^3)=\frac{2|H^{-1}|}{G}~,
\ee
where the first term on the left side is the mass of the over-dense region
and the second term is the mass outside the $R_J$ but inside the Hubble radius. 
As a result, we obtain $\delta_\text m=3\tilde \delta_\text m=9\left(\sqrt{\frac{2}{3}}\pi c_s \right)^{-3}$.

From Eq. (\ref{delta spectrum}), one can get the power spectrum of the density perturbation
\be
\Delta^2_{\delta}=\frac{k^3}{2\pi^2} |\delta_k|^2=\frac{k^4}{a^4 H^4} \Delta^2_{\mathcal R}~.
\ee
Accordingly, $\sigma$ is determined by
\be
\sigma^2=\int \frac{\text dk}{k} \Delta^2_{\delta}\ W\left(\frac{k R_\text J}{a}\right)=\int_{0}^{k_\text J} \frac{\text dk}{k} \Delta^2_{\delta}~,
\ee
where $W\big(\frac{k R_\text J}{a}\big)$ is the window function. For simplicity, we take
\[
W\left(\frac{k R_\text J}{a}\right)=
\begin{cases}
1 & k\leq k_\text J\\
0 & k> k_\text J
\end{cases} \ ,
\]
and then obtain
\be
\label{sigma}
\sigma=\frac{3}{4 c_s^2}\sqrt{\frac{a_{-}^3 H_{-}^2}{48\pi^2 c_s M_\text p^2 a^3}}~.
\ee

Therefore, the PBH mass fraction by (\ref{PS}) is
\begin{align}
\label{beta}
\beta(t)\simeq {\rm erfc} \bigg[ (48 c_s^{3/2}+9.355 c_s^{1/2}) \Big( \frac{M_\text p}{|H_{-}|} \Big) \Big( \frac{a(t)}{a_{-}} \Big)^{3/2} \bigg]  \nonumber \\
-{\rm erfc} \bigg[\frac{10.9427}{c_s^{1/2}} \Big( \frac{M_\text p}{|H_{-}|} \Big) \Big( \frac{a(t)}{a_{-}} \Big)^{3/2} \bigg]~,
\end{align}
for $c_s<0.39$; and $\beta(t)=0$ for  $c_s\geq 0.39$.
It can be seen that the value of $\beta$ increases as $a(t)$ becomes smaller along with the contracting and reaches its maximum value at the end of the matter contracting phase $t_{-}$.

\begin{figure}
\begin{center}
\includegraphics[width=0.7\textwidth]{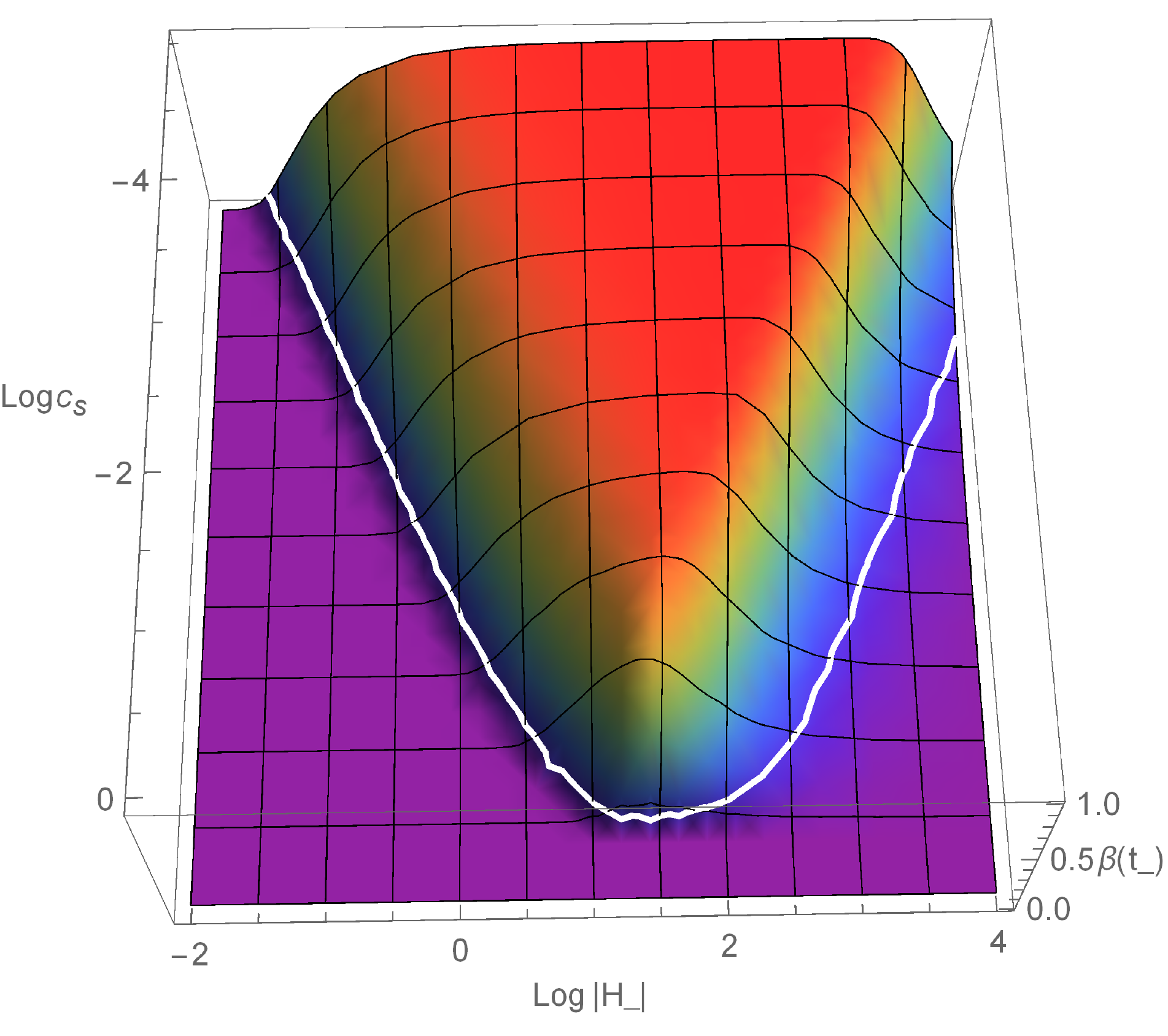}
\caption{\label{betaconstrain}$\beta(t_{-})$ varying from the model parameters $H_{-}$ and $c_s$.
The value of $\beta$ is illustrated by the color, decreasing from red to purple.
The isoline of $\beta(t_{-})=0.1$ is shown.}
\end{center}
\end{figure}

From the above expression, one can read that the maximal value of $\beta$ is unity, which corresponds to the case that the universe is dominated by black holes. Accordingly, we expect that $\beta\ll 1$ so that the universe is dominated by the background dust-like fluid. By numerically solving Eq. \eqref{beta}, we therefore derive the parameter space of $|H_{-}|$ and $c_s$ for each fixed value of $\beta$ as shown in Fig.~\ref{betaconstrain}.
From this figure, one can read that the non-vanishing mass fraction $\beta(t_{-})$, when $c_s<0.39$,
is small in both the very low energy regime $|H_{-}|\ll M_\text p$ and very high one $|H_{-}|\gg M_\text p$.
In the high energy regime, the density fluctuation is generally large $\sigma \gg 1$ according to Eq. \eqref{sigma},
but $\beta(t_{-})$ is small due to the constraint that PBHs ought to be within the cosmic apparent horizon.
The constraint of $\beta$ from today's observation could be loose. For instance, if one assumes that PBHs constitute the totality of dark matter in the universe, then the value of $\beta$ could be of order $O(0.1)$ at most which might have occurred at the moment of matter-radiation equality (see for example \cite{Bartolo:2016ami} for related analyses). Therefore, if one considers the case of $\beta \lesssim 0.1$ and $c_s=10^{-4}$ as a specific example, then Eq. \eqref{beta} yields $|H_{-}| \lesssim 10^{-1} M_\text p$ or $|H_{-}| \gtrsim 10^{4} M_\text p$.
Note that, $|H_{-}| \lesssim 10^{-1} M_\text p$ can be easily satisfied, but $|H_{-}| \gtrsim 10^{4} M_\text p$ is disfavored from model construction.


\section{Evolution of PBH in bouncing phase}\label{evolve}

It is important to point out that the subsequent evolutions of PBHs after their formation are not considered in the above section.
Associated with the evolutions, including the accretion and the evaporation due to Hawking radiation,
one can obtain constraints upon bouncing cosmology.
This is the main subject in this section.
For simplicity, we only consider the PBH evolutions during the bouncing phase.

\subsection{The growth of PBH mass}

At first, we calculate the growth of PBH by accreting the mass around in the contracting background.
For simplicity, we still assume that the space outside the BH horizon is unperturbed Friedmann universe and the matter which flows into the horizon is due to the cosmic contracting.
For a BH with mass $M$ and radius $R$, the mass increased equals to that flowing into the horizon, with the speed $|H|R$,
\be
\label{evolution}
\dot M=4\pi R^2 \bar\rho |H|R~.
\ee
We shall investigate the evolution in the contracting era of the bouncing phase $t_{-}<t<0$. During this stage, the Hubble length diverges quickly (see Fig.~\ref{comhubble}), and all PBHs can be treated as Schwarzschild BHs and $R=2GM$, according to the arguements in the last section. Therefore, Eq. (\ref{evolution}) reduces to
\be
\dot R=-3 R^3 H^3=-3 R^3 \Upsilon^3 t^3~,
\ee
and the solution is
\be
\label{growth}
R(t)=\sqrt{\frac{1}{\frac{1}{R^2(t_{-})}+\frac{3}{2}\Upsilon^3 (t^4-H^4_{-}/\Upsilon^4)}}~.
\ee
It is obvious that, the BH radius increases with $t$ approaching to the bouncing point. Eq. (\ref{growth}) also gives a constraint as follows,
\be
R^2(t_{-})\leq \frac{2\Upsilon}{3 H^4_{-}}~.
\ee
For the models violating the constraints above, the mass of PBH will grow to infinity before the bouncing point. According to (\ref{RRR}), one can read that $R(t_{-})\geq \sqrt{\frac{2}{3}}\pi c_s |H_{-}^{-1}|$, and hence, the above constraint yields explicitly that\footnote{We note that, in fact, if $c_s$ is not very small, the radius of BH suffers an increasing after the bouncing begins, but this effect has been ignored in the present consideration for simplicity.}
\be
\label{constrain}
\Upsilon \geq c_s^2 \pi^2 H^2_{-}~,
\ee
for $c_s\leq 0.39$.
Eq.~(\ref{constrain}) implies that the bouncing phase is expected to proceed rapidly so that the mass of PBHs would not become divergent and the universe is safe from collapsing into the black hole completely.

\subsection{Backreaction and theoretical constraints}

The bouncing phase is driven by the background fluid which violates NEC in the frame of general relativity.
From the Friedmann equations not considering PBHs, the density and pressure of the background fluid at the bouncing point are $\rho_{\text {bg}}=0$ and $p_{\text {bg}}=-2M_\text p^2 \Upsilon$ respectively. To include the contribution of backreaction from PBHs, the Friedmann equation at the bouncing point can be written as
\be
\label{backreaction}
\frac{\ddot a}{a}=-\frac{1}{6M_\text p^2}(\rho_\text{BR}-6M_\text p^2 \Upsilon)~,
\ee
where $\rho_\text{BR}=\rho_{\rm PBH}+\rho_{\gamma}+3p_{\gamma}$ is the effective density of the backreaction, $\rho_{\rm PBH}$ is the density of PBH, $\rho_\gamma$ and $p_{\gamma}$ are the density and pressure of Hawking radiation respectively. For simplicity, we assume all Hawking radiation particles are ultra-relativistic, which satisfy the relation $p_{\gamma} = \rho_\gamma/3$.
If the back reactions neutralize the negative pressure of background $\rho_\text{BR}-6M_\text p^2 \Upsilon>0$, the universe would not expand any more, as mentioned in Section \ref{model}.
In the following, we will constrain the model through PBHs.

Recall that the energy density of PBHs is
\be
\label{rho}
\rho_{\rm PBH}=\frac{\langle M \rangle}{L^3}=\frac{4\pi M_\text p^2 \langle R \rangle}{L^3}~,
\ee
which depends on the mean PBH mass $\langle M \rangle$ or the mean PBH radius $\langle R \rangle$, and the mean PBH separation $L$. At the beginning of the bouncing phase $t_{-}$,
the PBH density is known
\be
\label{rho-}
\rho_{\rm PBH}(t_{-})=\beta (\rho_{\text{PBH}}+\rho_{\text{bg}})=3M_\text p^2 H_{-}^2 \beta(t_{-})~,
\ee
according to Eq. \eqref{betadef}, where the evolution of the earlier formed PBHs is not considered.
If $\langle M \rangle$ at the moment $t_{-}$ is known, one can obtain the mean PBH separation $L(t_{-})$  from Eqs. \eqref{rho} and \eqref{rho-}.
To obtain $\langle M(t_{-}) \rangle$, one should first calculated the mass function of PBHs.
Here  we assume that all PBHs have the same mass, which is a simplification generally used \cite{Josan},
and all PBHs are of the Jeans scale $\langle R(t_{-})\rangle \simeq\sqrt{\frac{2}{3}}\pi c_s |H_{-}^{-1}|$.
Therefore, the mean separation at $t_{-}$ is
\be
L(t_{-})=\frac{R(t_{-})}{\left(c_s^2\beta\right)^{\frac{1}{3}}}~.
\ee
At the present investigation, we have neglected the newly formed PBHs during the bouncing phase.
Therefore, the mean separation follows $L(t) \propto a(t)$, and accordingly, $L(0) \simeq L(t_{-}) e^{-\frac{H_{-}^2}{2\Upsilon}}$. Note that the PBH evaporation due to Hawking radiation has been studied clearly in \cite{Carr:2009jm}. Following this, one can estimate the lifetime of PBH to be
\be
\tau \simeq 13.7{\rm Gyr} \left( \frac{M}{5\times 10^{14} {\rm g}}\right)^3 \simeq 5\times10^3 \frac{M^3}{M_\text p^4}~.
\ee

We have adopted the approximation that for massive PBHs $\tau > |t_{-}|$, or $\Upsilon > {H_{-}^4}/{(10^8 c_s^3 M_\text p^2)}$, the Hawking radiation is sub-dominant. A closer analysis combing both the Hawking radiation and the growth of PBHs will be addressed in our follow-up study.
As a result, the evolution of the PBHs follows Eq. (\ref{growth}), and their size approximately takes $R(0) \simeq R(t_{-})\left( \frac{\Upsilon}{\Upsilon-c_s^2 \pi^2 H^2_{-}}\right)^{1/2}$ at the bouncing point. Otherwise, PBH would have evaporated completely before the universe reaches the bouncing point with $R(0)=0$. As a result, we can estimate the energy density of PBHs at the bouncing point as
\be
\label{rhoPBH}
\rho_{\rm PBH}(0)=\frac{4\pi M_\text p^2 R(0)}{L(0)^3}=
 \begin{cases}
\frac{6M_\text p^2 H_{-}^2 }{\pi} \beta(t_{-}) \sqrt{\frac{\Upsilon}{\Upsilon-c_s^2\pi^2H_{-}^2}} \  e^{\frac{3H_{-}^2}{2\Upsilon}}, &  \Upsilon > \frac{H_{-}^4}{10^8 c_s^3 M_\text p^2}\\
0, & \Upsilon < \frac{H_{-}^4}{10^8 c_s^3 M_\text p^2}
 \end{cases}~.
\ee
Similarly, we analyze the backreaction due to the Hawking radiation. For $\Upsilon > \frac{H_{-}^4}{10^8 c_s^3 M_\text p^2}$, the Hawking radiation can be ignored and $\rho_{\gamma}(0)=0$. In the limit that PBHs finish their evaporation at $t_{-}$, one has $\rho_{\gamma}(t_{-})=\rho_\text{PBH}(t_{-})$ and then $\rho_{\gamma} \propto a^{-4}$. In the limit that the effective evaporation happens at the bouncing point, $\rho_\gamma(0)=\rho_\text{PBH}(0)$, and $\rho_\text{PBH} \propto a^{-3}$ before its evaporation. Therefore, one has
\be
\label{rhogamma}
\rho_{\gamma}(0)=
 \begin{cases}
0 ~, & \Upsilon > \frac{H_{-}^4}{10^8 c_s^3 M_\text p^2} \\
\rho_\text{PBH}(t_{-}) \  \Big( e^{\frac{H_{-}^2}{2\Upsilon}} \Big)^n ~, & \Upsilon < \frac{H_{-}^4}{10^8 c_s^3 M_\text p^2}
 \end{cases} ~,
\ee
where $\rho_{\text{PBH}}(t_{-})$ is given by Eq. (\ref{rho-}) and $n$ is a parameter that describes the moment of the complete evaporation of PBH. For instance, $n=4$ corresponds to the limit that the evaporation happens at $t_{-}$; and, $n=3$ corresponds to this limit at the bouncing point.

\begin{figure}
\begin{center}
\includegraphics[width=0.6\textwidth]{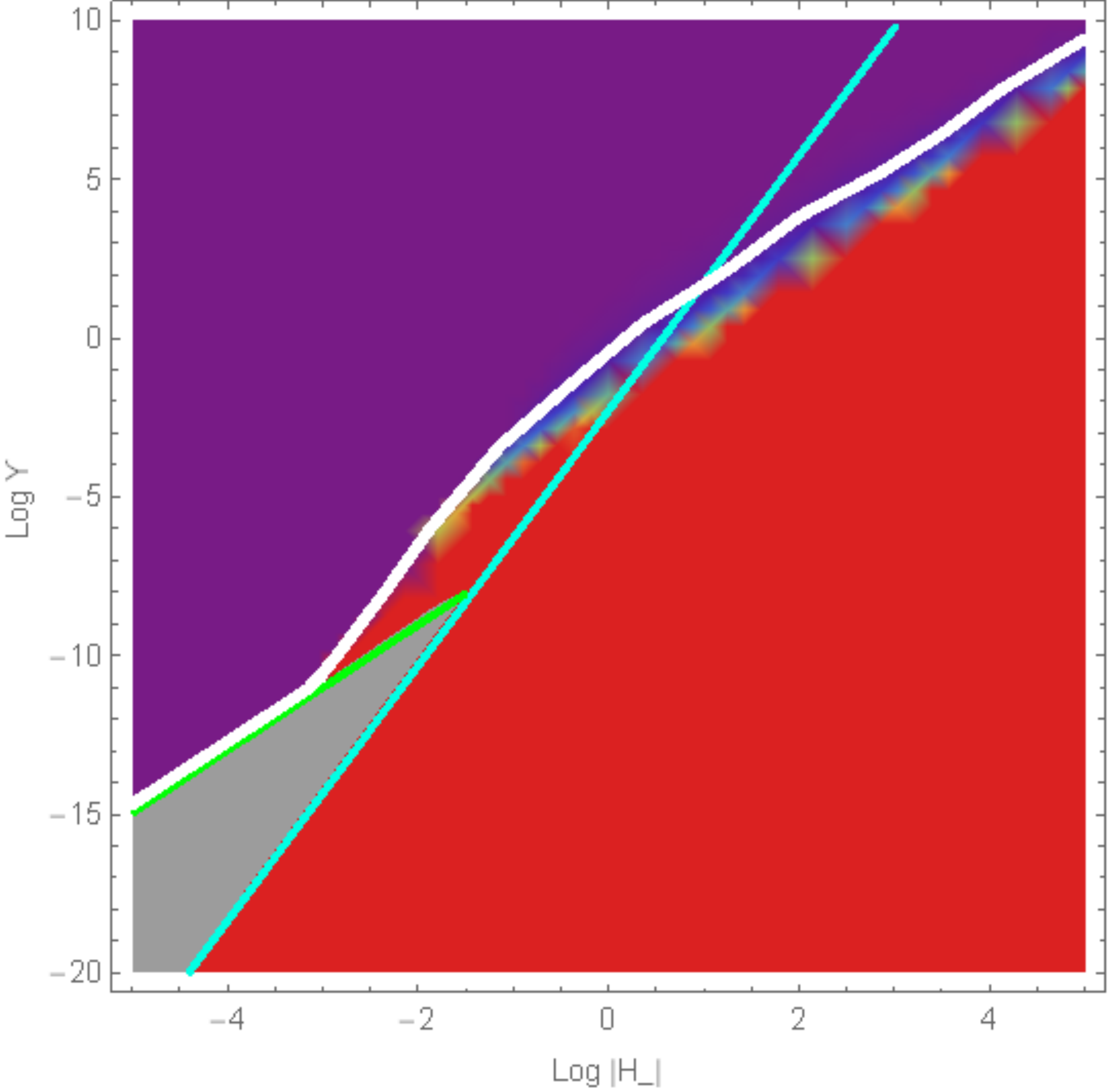}
\caption{\label{cs=-3} The back reaction $\rho_\text{BR}/|\bar p|$ varying from model parameters $H_{-}$ and $\Upsilon$, with  $c_s=10^{-3}$  and $n=4$ taken. The value of $\rho_\text{BR}/|\bar p|$ is illustrated by color, increasing from purple to red.
The white thick curve is the isoline $\rho_\text{BR}/|\bar p|=1$, and only the parameter space over this curve is allowed.
The light blue curve is the boundary between PBH growth and evaporation $\Upsilon = \frac{H_{-}^4}{10^8 c_s^3 M_\text p^2}$, the area over the boundary is the PBH growing region, and that under the boundary is the  evaporation region. The gray region is the forbidden parameter space due to Eq. (\ref{constrain}), and the upper boundary of this region is $\Upsilon = c_s^2 \pi^2 H^2_{-}$ illustrated by green curve. }
\end{center}
\end{figure}

Inserting Eqs. (\ref{rhoPBH}) and (\ref{rhogamma}) into the Friedmann equation (\ref{backreaction}), in order not to neutralize the negative pressure $\bar p = -2 M_\text p^2 \Upsilon$, the following constraint should be satisfied
\be
\label{negativeconstrain}
 \frac{\rho_\text{BR}}{|\bar p|}=
 \begin{cases}
\frac{\beta(t_{-}) H_{-}^2}{\pi \Upsilon} \sqrt{\frac{\Upsilon}{\Upsilon-c_s^2\pi^2H_{-}^2}} \ e^{\frac{3H_{-}^2}{2\Upsilon}} < 1, &  \Upsilon > \frac{H_{-}^4}{10^8 c_s^3 M_\text p^2} \\
\frac{\beta(t_{-}) H_{-}^2}{\Upsilon} \Big( e^{\frac{H_{-}^2}{2\Upsilon}} \Big)^n < 1,  & \Upsilon < \frac{H_{-}^4}{10^8 c_s^3 M_\text p^2}
 \end{cases}~.
\ee
The resulting constraints are numerically shown in Fig. \ref{cs=-3} for the model with $c_s=10^{-3}$. It is seen that the lines of $\Upsilon = c_s^2 \pi^2 H^2_{-}$ and $\Upsilon=\frac{H_{-}^4}{10^8 c_s^3 M_\text p^2}$, which represent the limit case of (\ref{constrain}) and  the boundary between the PBH growth and evaporation respectively, play important roles in the constraints. The two lines shape the forbidden parameter space in which PBH grows into infinity. Furthermore, two lines intersect at the characteristic energy scale $H_{-}^2 \simeq 10^9 c_s^5 M_\text p^2$. At the low energy $H_{-}^2 \ll 10^9 c_s^5 M_\text p^2$, $\Upsilon = c_s^2 \pi^2 H^2_{-}$ is the asymptote for all isolines of $\rho_\text{BR}/|\bar p|$. Therefore, the constraint $\rho_\text{BR}/|\bar p|<1$ reduces to Eq. (\ref{constrain}) at the low energy.

\subsection{Estimates of observational constraints}

After the bouncing phase, PBHs will evolve in the regular expanding universe, and hence can be constrained by today's observations.
In this subsection, we briefly estimate the constraints upon the bouncing models though the observational limits of PBHs.

For massive PBHs with an initial mass $M>10^{15}$g, they could survive until today and their density scales as $a^{-3}$ approximately in the expanding universe\cite{Carr:2009jm}. As a result, the current PBH($M>10^{15}$g) density parameter can be expressed as
\[
\Omega_{\rm PBH}\simeq\frac{\rho_{\rm PBH}(0)}{3 M_\text p^2 H_0^2} \left(\frac{a_\text b}{a_0} \right)^3,
\]
where $a_0$ and $H_0$ denote the scale factor and Hubble parameter of today, $a_\text b$ is the scale of bounce, and $\rho_{\rm PBH}(0)$ is the PBH density at the bouncing point.
Since the PBHs may contribute to part of dark matter, one immediately has the rough constraint $\Omega_{\rm PBH}<\Omega_{\rm DM}\simeq0.25$ and thus can get a bound
\be
\label{OmegaPBH1}
\rho_{\rm PBH}(0)<3 M_\text p^2 H_0^2 \Omega_{\rm DM} \left(\frac{a_\text b}{a_0} \right)^{-3}.
\ee
For the light PBHs with $M<10^{15}$g, they have been completely evaporated today.
Therefore, the density parameter of the Hawking radiation $\Omega_{\rm HR}$ is approximately estimated as
\be
\Omega_{\rm HR}\simeq\frac{\rho_{\rm PBH}(0)}{3 M_\text p^2 H_0^2} \left(\frac{a_\text r}{a_0} \right)^4 \left(\frac{a_\text b}{a_\text r} \right)^3 >\frac{\rho_{\rm PBH}(0)}{3 M_\text p^2 H_0^2} \left(\frac{a_\text b}{a_0} \right)^4,
\ee
where $a_\text r$ is the scale factor when PBHs complete the evaporation, and $a_\text b<a_\text r<a_0$.
\begin{figure}
\begin{center}
\includegraphics[width=0.7\textwidth]{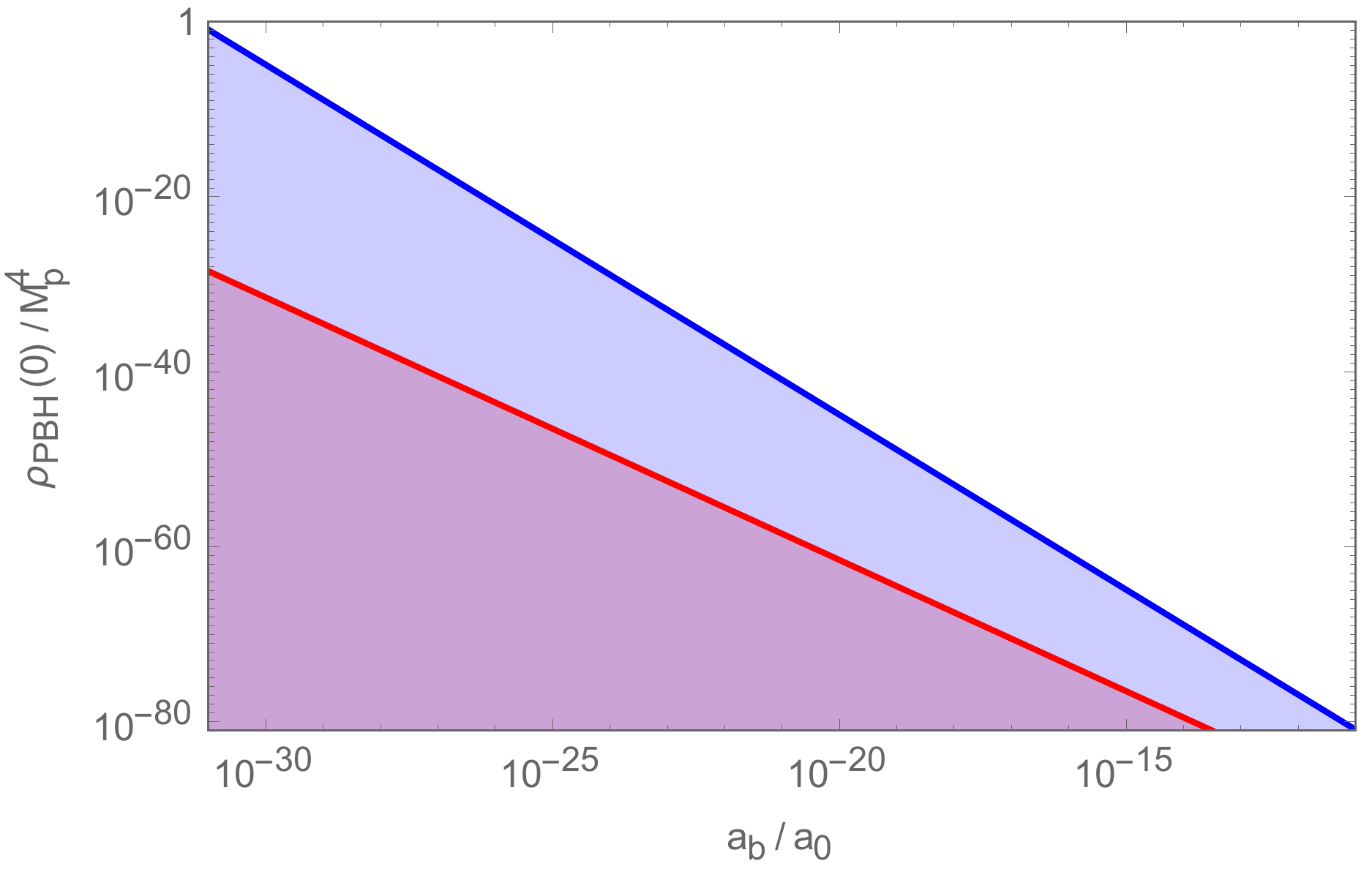}
\caption{\label{Omega}
The constraints on $\rho_{\text{PBH}}(0)$ along with $a_b/a_0$.
The blue shadowed region is given by (\ref{OmegaPBH1}) for PBHs with $M>10^{15}$g, and the red shadowed region is given by (\ref{OmegaPBH2}) for $M<10^{15}$g.}
\end{center}
\end{figure}
$\Omega_{\rm HR}$ is not allowed to be larger than the total density of today's radiation $\Omega_{\rm r}\sim 10^{-4}$, measured by the CMB experiments.
As a consequence, one gets
\be
\label{OmegaPBH2}
\rho_{\rm PBH}(0)<3 M_\text p^2 H_0^2 \Omega_{\rm r} \left(\frac{a_\text b}{a_0} \right)^{-4}.
\ee
In Fig. \ref{Omega}, we provide the constraints on $\rho_{\text{PBH}}(0)$ along with $a_b/a_0$ from the bounds \eqref{OmegaPBH1} and \eqref{OmegaPBH2}, respectively. The PBH mass $M$ and density $\rho_{\rm PBH}(0)$ are functions of $c_s$, $H_{-}$ and $\Upsilon$, which are provided in the previous subsection. As a result, we find that the nonsingular bounce models can be constrained by observations due to \eqref{OmegaPBH1} and \eqref{OmegaPBH2}.
We note that these constraints are quite loose. More stringent constraints can be obtained by taking the observational bounds of $\Omega_{\text{PBH}}$ and $\Omega_{\text{HR}}$ from the CMB, gamma-ray burst and BBN \emph{etc.} \cite{Carr:2009jm}.

\section{Conclusion and outlook}\label{conclusion}

In this paper, we have investigated the formation and evolution of PBHs in the matter bounce cosmology.
Firstly, we described the general matter bounce models by some parameters like $c_s$, $H_{-}$ and $\Upsilon$.
The comoving curvature perturbation $\mathcal R_k$ is also calculated during the matter contracting phase, which seeds the PBH formation.
Then we had a discussion about the condition of PBH formation in the contracting background, which is different from that in the expanding universe.
By taking a simple collapsing model, the threshold of the density fluctuation for forming a PBH is derived.
Furthermore, in the comoving gauge, the density fluctuation and its threshold are all related to the curvature perturbation $\mathcal R_k$.
Therefore, we can calculate the mass fraction of PBHs $\beta$ in the contracting phase from the Press-Schechter theory,
and constrain the bouncing models from PBHs.
PBH formation depends on the model parameters $c_s$ and $H_{-}$.
For instance, PBHs can form in the contracting phase only if $c_s\leq 0.39$,
since the BHs are not allowed to be larger than the cosmic apparent horizon.
When $c_s\leq 0.39$, $\beta$  is small and the model is safe for  $|H_{-}|\ll M_\text p$ or $|H_{-}|\gg M_\text p$,
but only the low energy regime is favored from the model construction.

The subsequent evolution of  PBHs in the bouncing phase is also investigated, with the PBH accretion and  Hawking radiation considered. The growth behavior of PBH yields a constraint to the model $\Upsilon \geq c_s^2 \pi^2 H^2_{-}$, in case that the mass of PBHs grow to infinity before the bouncing point. Moreover, the back reaction of PBH and its Hawking radiation is calculated to constrain models, in order not to neutralize the negative pressure $p_{\rm bg}=-2M_\text p^2 \Upsilon$, since in such case the universe cannot expand again. The constraint reduces to $\Upsilon \geq c_s^2 \pi^2 H^2_{-}$ at the low energy scale $H^2_{-}\ll 10^9 c_s^5 M_\text p^2$.
Afterwards, a rough constraint of bouncing model though the PBH observations is given, and the constraint is stringent only when $\Upsilon$ is slightly larger than $c_s^2 \pi^2 H^2_{-}$.
A more precise analysis taken into account both the PBH growth and the Hawking evaporation as well as the detailed constraints from cosmological observations will be addressed in our follow-up works.

We note that while our paper was being prepared, an independent work was being carried out by another group \cite{Quintin:2016qro}, which explores similar features of BH formation in bouncing cosmology.

\section*{Acknowledgments}
We are grateful to R. Brandenberger, J. Martin, P. Peter, T. Qiu, J. Quintin, D.-G. Wang, Z. Wang, X.-Y. Yang and Y. Zhang for valuable comments. We especially thank the anonymous Referee for very detailed and constructive suggestions.
YFC and JWC are supported in part by the Chinese National Youth Thousand Talents Program (Grant No. KJ2030220006), by the USTC start-up funding (Grant No. KY2030000049) and by the National Natural Science Foundation of China (Grant Nos. 11421303, 11653002).
HLX is supported in part by the Fund for Fostering Talents in Basic Science of the National Natural Science Foundation of China (Grant No. J1310021) and in part by the Yan Jici Talent Students Program.
HLX also thanks J. Martin and P. Peter for their warmest hospitality during visiting the Institut d'Astrophysique de Paris when this work was finalized.
Part of numerical computations are operated on the computer cluster LINDA in the particle cosmology group at USTC.

\end{document}